\documentclass[twocolumn,showpacs,preprintnumbers,amsmath,amssymb]{revtex4}
\usepackage{graphicx}

\begin{document}
\title{Practically secure quantum bit commitment based on quantum seals}
\author{Guang Ping He}
 \email{hegp@mail.sysu.edu.cn}
\affiliation{School of Physics \& Engineering and Advanced
Research Center, Sun Yat-sen University, Guangzhou 510275, China\\
and Center of Theoretical and Computational Physics, The University
of Hong Kong, Pokfulam Road, Hong Kong, China}
\author{Z. D. Wang}
 \email{zwang@hkucc.hku.hk}
\affiliation{Department of Physics and Center of Theoretical and
Computational Physics, The University of Hong Kong, Pokfulam Road,
Hong Kong, China}

\begin{abstract}
The relationship between the quantum bit commitment (QBC) and
quantum seal (QS) is studied. It is elaborated that QBC and QS are
not equivalent, but QS protocols satisfying a stronger unconditional
security requirement can lead to an unconditionally secure QBC. In
this sense, QS is strictly stronger than QBC in security
requirements. Based on an earlier proposal on sealing a single bit,
a feasible QBC protocol is also put forward, which is secure in
practice even if the cheater has a strong quantum computational
power.
\end{abstract}


\maketitle

\newpage

\section{INTRODUCTION}

Bit commitment (BC) is a cryptographic task between two participants
Alice and Bob. It normally includes two phases. In the commit phase,
Alice has in mind a bit ($b=0$ or $1$) which she wants to commit to
Bob, and she sends him a piece of evidence. Later, in the unveil
phase, Alice announces the value of $b$, and Bob checks it with the
evidence. A BC protocol is said to be binding if Alice cannot change
the value of $b$ after the commit phase, and is said to be
concealing if Bob cannot know $b$ before the unveil phase. A secure
BC protocol needs to be both binding and concealing. Quantum bit
commitment (QBC) \cite{BCJL93} is known to be an essential element
for building more complicated ``post-cold-war era'' multi-party
cryptographic protocols, e.g., quantum coin tossing and quantum
oblivious transfer \cite {Yao,Kilian}. Unfortunately, according to
the Mayers-Lo-Chau (MLC) no-go theorem \cite{Mayers,LC},
unconditionally secure QBC does not seem to exist in principle.  On
the other hand, quantum seals (QS) are relatively new to quantum
cryptography. Its goal can be summarized as follows. Alice, the
owner of the secret data to be sealed, encodes the data with quantum
states. Any reader Bob can obtain the data from these states without
the help of Alice, while reading the data should be detectable by
Alice. A QS protocol is considered to be secure if it possesses the
following feature: Bob cannot escape Alice's detection once he reads
the data, namely, Bob is unable to read the data if he wants to
escape Alice's detection. QS can be classified by the types and the
readability of the sealed data. If the data is a classical bit, it
is called quantum bit seals (QBS). Else if the data is a classical
string, it is called quantum string seals (QSS). If the data can
always be obtained by the reader with certainty, it is called a
perfect QS. Else if the data can only be obtained with a small but
non-vanished error rate, it is an imperfect QS. The first perfect
QBS protocol was proposed by Bechmann-Pasquinucci \cite{sealing},
but then it was proven that all perfect QBS ones are insecure
against collective measurements \cite{impossibility}. Shortly later,
it was also found that imperfect QBS  has security limits \cite{He}.
Nevertheless, the imperfect QSS can be secure
\cite{String,Insecure,note,Security,qi505,qi516}.   More
intriguingly, it was proposed in Ref. \cite{String} that secure QSS
can be utilized to realize a kind of QBS, which is secure in
practice.

Though QBC and QS may look like irrelevant at the first glance, here
we elaborate that they are closely related. More rigorously, it will
be shown below that a QBS protocol satisfying certain security
conditions can lead to an unconditionally secure QBC. Based on the
method of realizing QBS with QSS in Ref. \cite {String}, a new QBC
protocol is obtained. A most intriguing feature of this QBC protocol
is that the present QBS protocol enables the secret data to be
decoded with common senses of human being, thus it is very easy for
an honest participant. On the contrary, a dishonest participant
cannot fulfill the coherent attacks even if he has a very powerful
quantum computer that can perform any kind of quantum operations and
measurements. Instead, he  needs also a knowledge base that has
almost all knowledge of human culture, and his quantum computer
needs to have a high level of artificial intelligence to handle
these knowledge. This is impossible not only for current technology,
but also is less likely in practice even in the future, because
human intelligence seems to contain non-computable elements
\cite{Penrose}. In this sense, the present QBC protocol may be
viewed as ``practically secure''. Moreover, this QBC is quite
feasible. Thus it is possible to build more complicated QBC-based
multi-party cryptographic protocols and to enjoy the advantage of
quantum cryptography in practice, even though the MLC no-go theorem
is still there.

\section{The relationship between quantum bit commitment and quantum seals}

For a QBS protocol with a known error rate $\varepsilon $, we can
build a QBC protocol as follows (note that in this section, the
names Alice and Bob are called contrarily to those in the
description of QS in the Introduction, because the QBS protocol is
executed in the reverse direction in the QBC protocols below).

\bigskip

\textit{The Basic Protocol}

\bigskip

The commit phase:

(1) Bob randomly chooses $s$\ bits $x_{i}$ ($i=1,...,s$), and seals each bit
into a quantum register $\psi _{i}$ ($i=1,...,s$) with the QBS protocol.
Then he sends these quantum registers to Alice.

(2) Alice randomly chooses $m$ of the quantum registers and decodes
their corresponding sealed bits. She checks if the values of these
$m$ sealed bits are random. She also asks Bob to reveal the values
of the sealed bits for these $m$ quantum registers to check if they
match the values she decoded within the error rate $\varepsilon $.
If no suspicious result was found, Alice and Bob discard the data
for these $m$ quantum registers and proceed.

(3) Alice randomly chooses one of the remaining $s-m$ quantum registers
(suppose that it is $\psi _{i_{0}}$), decodes the sealed bit $x_{i_{0}}$,
and leaves the rest $s-m-1$ quantum registers unmeasured. Now she has in
mind the value of the bit $b$ she wants to commit, and announces to Bob the
value of $a=x_{i_{0}}\oplus b$.

\bigskip

The unveil phase:

(4) Alice announces to Bob the value of the commit bit $b$ and $i_{0}$, and
sends Bob all the $s-m-1$ unmeasured quantum registers.

(5) Bob uses the value $x_{i_{0}}$ he sealed in $\psi _{i_{0}}$ to check whether $%
a=x_{i_{0}}\oplus b$, and he checks whether the $s-m-1$ quantum
registers Alice returns him in this phase are indeed unmeasured,
i.e., the corresponding sealed bits have not been decoded. He
accepts Alice's commitment as honest if no suspicious result was
found.

\bigskip

This protocol is concealing because Bob cannot know $i_{0}$ before
the unveil phase, since Alice's announcing $a$ provides him $1$ bit
of information only, while locating $i_{0}$ requires $\log
_{2}(s-m-1)$ bits of information. Therefore Bob does not know
$x_{i_{0}}$ so he cannot infer the
value of the commit bit $b$ from Alice's announced $a$, as long as each $%
x_{i}$\ has the equal probabilities to be $0$ or $1$ (that is why Alice
needs to check in step (2) whether the values of the sealed bits are
random). It is also binding because if Alice wants to change the value of $b$
in the unveil phase, she has to find another quantum register $\psi _{i_{1}}$%
\ whose sealed bit is $x_{i_{1}}=\bar{x}_{i_{0}}$, and sends $\psi _{i_{0}}$
to Bob as an unmeasured quantum register in step (5). But if the QBS
protocol used in this QBC protocol is secure, Bob will detect that the bit
sealed in $\psi _{i_{0}}$ has already been decoded, and catch Alice cheating.

However, Alice can apply the following weak cheating strategy. She measures
no register in step (3), and announces a random value as $a$. Then in the
unveil phase, she randomly chooses a register as $\psi _{i_{0}}$, and
determines what value of $b$ can be announced in step (4) according to the
decoding result of her current measurement on $\psi _{i_{0}}$ and the value
of $a$ she previously announced. This may not be considered as a ``real''
cheating because Alice cannot announce $b$ at her will in the unveil phase.
It is like tossing a coin and decide the value of $b$ by the head or tail.
But on the other hand, it is also true that the protocol cannot limit Alice
to honest behaviors because she is not forced to make up her mind on the
value of $b$ during the commit phase. To fix this problem, we can further
improve the protocol.

\bigskip

\textit{The Advanced Protocol}

\bigskip

The commit phase:

(i) Alice and Bob agree on the security parameters $s$, $m$ and $n$\ ($%
n<<s-m $), and a Boolean matrix $G$\ as the generating matrix of a binary
linear $(n,k,d)$-code $C$\ \cite{code}.

(ii) The same as steps (1) and (2) of the basic protocol.

(iii) Alice randomly chooses $n$ of the remaining $s-m$ quantum registers.
Suppose that they are $\psi _{j}$ ($j=i_{1},i_{2},...,i_{n}$, $%
i_{1}<i_{2}<...<i_{n}$). She decodes each of the bit $x_{j}$ sealed in these
$n$ registers, and leaves the rest $s-m-n$ quantum registers unmeasured.
Thus she obtains a classical $n$-bit string $%
x=(x_{i_{1}}x_{i_{2}}...x_{i_{n}})$.

(iv) Alice chooses a non-zero random $n$-bit string $r=(r_{1}r_{2}...r_{n})%
\in \{0,1\}^{n}$ and announces it to Bob;

(v) Now Alice has in mind the value of the bit $b$ that she wants to commit.
She chooses an $n$-bit codeword $c=(c_{1}c_{2}...c_{n})$ from $C$ such that $%
c\odot r=b$ (Here $c\odot r\equiv \bigoplus\limits_{i=1}^{n}c_{i}\wedge
r_{i} $);

(vi) Alice announces to Bob $c^{\prime }=c\oplus x$.

\bigskip

The unveil phase:

(vii) Alice announces to Bob the value of the commit bit $b$ and $%
i_{1},i_{2},...,i_{n}$, and sends Bob all the $s-m-n$ unmeasured quantum
registers.

(viii) Bob uses the values $x_{j}$ he sealed in $\psi _{j}$ ($%
j=i_{1},i_{2},...,i_{n}$) to obtain $c$\ from $c^{\prime }=c\oplus x$, and
checks whether $c$ is indeed a codeword from $C$ and $c\odot r=b$. He also
checks whether the $s-m-n$ quantum registers Alice sends him in this phase
are indeed unmeasured, and accepts Alice's commitment as honest if no
suspicious result was found.

\bigskip

The intuition behind this protocol is that if Alice wants to keep the value
of $b$ undetermined in the commit phase by measuring less than $n$ registers
in step (iii), she has to announce $c^{\prime }$ randomly in step (vi) since
she does not know $x$. Then in step (vii) of the unveil phase, she has to
find the correct indices $i_{1},i_{2},...,i_{n}$ to announce, so that the
data $x=(x_{i_{1}}x_{i_{2}}...x_{i_{n}})$ sealed in $\psi _{j}$ ($%
j=i_{1},i_{2},...,i_{n}$) ensures that $c^{\prime }\oplus x$ is a
codeword from $C$. Generally, if she randomly chooses $n$ indices to
announce while leaving the rest $s-m-n$ quantum registers completely
unmeasured, the probability is trivial for the value of $x$ thus
obtained to be exactly what needed. The question is whether the
following strategy can be successful: In the unveil phase, Alice
randomly chooses $n$ quantum registers and measures them with
collective measurements instead of individual measurements, so that
they can be less disturbed if their sealed data is not the desired
$x$. Then Alice can repeat the procedure with other sets of $n$
registers, until she finally finds the set in which the desired $x$
is sealed. In this case, whether the above QBC protocol is secure or
not will be determined by how much disturbance is caused by Alice's
collective measurements. That is, it is determined by the security
level of the QBS protocol used in step (ii). To find out how secure
this QBS needs to be, let us first review the classification of
definitions of security of cryptographic protocols. Currently, the
following two definitions of security are widely adopted \cite
{Mayers}.

\bigskip

\textit{Perfect security:}

When the other participant(s) is (are) honest, the amount of extra
information (other than what is allowed by the protocol) obtained by the
dishonest participant(s) is exactly zero.

\bigskip

\textit{Unconditional security:}

When the other participant(s) is (are) honest, the amount of extra
information obtained by the dishonest participant(s) can be made arbitrarily
small by increasing the security parameter(s) of the protocol.

\bigskip

Obviously when perfectly secure QBS protocols are used in Advanced Protocol,
it will result in an unconditionally secure QBC. What we are interested is
whether unconditionally secure QBS protocols are useful too. Note that the
above definition of unconditional security is equivalent to: when the other
participant(s) is (are) honest, the probability for a dishonest participant
to obtain non-trivial amount of extra information while escaping the
detection can be made arbitrarily small by increasing the security
parameter(s) of the protocol. However, it was left blank in this definition
whether the probability for a dishonest participant to escape the detection
can be made arbitrarily small if he obtains trivial amount of extra
information only. To be rigorous, here we propose an independent definition
for a kind of unconditionally secure protocols which satisfies stricter
security requirement.

\bigskip

\textit{Strong unconditional security:}

When the amount of extra information obtained by a dishonest participant is $%
f(n)$ (here $f(n)$\ is a function whose value decreases as the security
parameter(s) $n$ of the protocol increases), the probability for him to
escape the detection by the honest participant(s) is $o(\alpha ^{f(n)})$.
Here the constant $\alpha $ satisfies $0<\alpha <1$.

\bigskip

Now let us see what difference will be made when different kinds of
unconditionally secure QBS protocols are used in the above Advanced
QBC Protocol. Suppose that Alice applies the cheating strategy
mentioned above. She measures no quantum registers in the commit
phase. In the unveil phase, she randomly chooses $n$ registers and
measures them with collective measurements. If it turns out that the
data sealed in these registers is exactly the desired $x$ that could
lead to the value of $b$ she wants to unveil, then her cheating is
successful. Else she tries again with another set of $n$ registers.
Note that in the latter case, she merely knows that the data sealed
in the first $n$ registers she measured is not $x$, while she does
not know what exactly the sealed data is. Thus the amount of
information she obtained is merely $f(n)$. If general
unconditionally secure QBS protocols are adopted which do not
satisfy the strong unconditional security condition, Alice can
escape the detection with a probability not less than $O(\alpha
^{f(n)})$. Suppose that it averagely takes $t$ times for Alice to
perform collective measurements on different sets of $n$ registers
until she finds the desired $n$ registers which give the outcome $x$
she is looking for. Since the goal of Alice is to commit $1$ bit of
information, we have $tf(n)\sim O(1)$, i.e., $tf(n)$ has a finite
non-trivial value. Then the probability for Alice to escape the
detection will be not less than $[O(\alpha ^{f(n)})]^{t}\sim
O(\alpha ^{tf(n)})$, which also remains finite instead of being
trivial as $n$ increases. Thus we come to the conclusion that
\textit{unconditionally secure QBS which does not satisfy the strong
unconditional security condition does not necessarily lead to
unconditionally secure QBC} through the above Advance Protocol.

On the other hand, if strong unconditionally secure QS protocols are
adopted in Advanced Protocol, whenever Alice measures $n$ registers
and obtains $f(n)$ bits of information, she can escape the detection
with probability $o(\alpha ^{f(n)})$ only. That is, when she finally
finds the $n$ registers she is looking for and thus obtains finite
non-trivial amount of information, the probability for her to escape
the detection will be $\sim o(\alpha ^{tf(n)})$, which does not
remain finite. Instead, it drops arbitrarily close to zero as $n$
increases. Consequently, the resultant QBC protocol is
unconditionally secure. Therefore we reach one of the main result of
this paper: \textit{strong unconditionally secure QBS can lead to
unconditionally secure QBC}.

It is natural to ask a further question whether strong unconditionally
secure QBS and unconditionally secure QBC are equivalent. So far, to our
best knowledge, no unconditionally secure QBS protocol based on
unconditionally secure QBC (no matter it exists or not) was ever proposed.
It is also worth noting that the definitions of QBS and QBC indicate a
significant different feature between them. In the final stage of QBC, Alice
needs to unveil to Bob what she committed, i.e., classical communication is
needed. On the contrary, in QBS Alice can check whether the sealed data was
decoded or not at any time, with or without the classical communication from
Bob. Therefore it seems reasonable that an unconditionally secure QBS
protocol based on unconditionally secure QBC does not exist at all. For
these reasons, we tend to believe that strong unconditionally secure QBS and
unconditionally secure QBC are not equivalent. \textit{Strong
unconditionally secure QBS is strictly stronger than unconditionally secure
QBC in security requirements}.

\section{Practically secure quantum bit commitment protocol}

In Ref. \cite{He}, a model of QBS was proposed and it was proven that any
QBS satisfying this model cannot be unconditionally secure. To date, the
model seems to cover all QBS protocols within our current imagination.
Consequently, before we can come up with a protocol which satisfies the
definition of QBS exactly while is not covered by that model,
unconditionally secure QBS does not seems to exist in principle. Therefore
no QBS can lead to unconditionally secure QBC via the above Advanced
Protocol. Nevertheless, it was shown in Ref. \cite{String} that
unconditionally secure QSS exits, and it can lead to QBS which is secure in
practice. Therefore we can use this QBS to realize a kind of QBC which is
also secure in practice.

Let us briefly review the QSS and QBS protocols proposed in Ref.
\cite {String}. Let $\Theta $\ ($0<\Theta \ll \pi /4$) and$\ \alpha
$\ ($0<\alpha <1/2$) be two fixed constants. The proposed
\textit{QSS protocol} is as follows.

\bigskip

\textit{Sealing}: To seal a classical $N$-bit string $b=b_{1}b_{2}...b_{N}$ (%
$b_{i}\in \{0,1\}$), Alice randomly chooses $\theta _{i}$ ($-\Theta
/N^{\alpha }\leqslant \theta _{i}\leqslant \Theta /N^{\alpha }$) and encodes
each bit $b_{i}$ with a qubit in the state $\left| \psi _{i}\right\rangle
=\cos \theta _{i}\left| b_{i}\right\rangle +\sin \theta _{i}\left| \bar{b}%
_{i}\right\rangle $. She makes these $n$ qubits publicly accessible to the
reader, while keeping all $\theta _{i}$ ($i=1,...,N$) secret.

\textit{Reading}: When Bob wants to read the string $b$, he simply measures
each qubit in the computational basis $\{\left| 0\right\rangle ,\left|
1\right\rangle \}$, and denotes the outcome as $\left| b_{i}^{\prime
}\right\rangle $. He takes the string $b^{\prime }=b_{1}^{\prime
}b_{2}^{\prime }...b_{N}^{\prime }$ as $b$.

\textit{Checking}: At any time, Alice can check whether the sealed string $b$
has been read by trying to project the $i$-th qubit ($i=1,...,N$) into $\cos
\theta _{i}\left| b_{i}\right\rangle +\sin \theta _{i}\left| \bar{b}%
_{i}\right\rangle $. If all the $N$ qubits can be projected successfully,
she concludes that the string $b$ is still unread. Otherwise if any of the
qubits fails, she knows that $b$ is read.

\bigskip

This protocol can achieve the following goal: each bit sealed by Alice can
be read correctly by Bob, except with a probability not greater than $%
\varepsilon \equiv \sin ^{2}(\Theta /N^{\alpha })$. Thus by
increasing $N$, the maximal reading error rate $\varepsilon $\ can
be made arbitrarily small. Meanwhile, the total probability for Bob
to obtain $K$ bits of information while escaping the detection is
bounded by
\begin{equation}
P\leqslant 2^{-K}\prod\limits_{i=1}^{N}2\cos ^{2}\theta _{i},
\end{equation}
which drops exponentially as $K\rightarrow N$, and vanishes when $%
N\rightarrow \infty $ as long as $0<\alpha <1/2$. Thus the protocol is
unconditionally secure in principle.

This QSS protocol can be utilized to seal a single bit. The method is to
turn the sealed bit into a string by adding redundant information. There are
many ways to accomplish this. In the simplest case, we can map some global
properties of the string, e.g., the parity or the weight, into the bit we
want to seal. But since such kinds of mapping is easy to be formulated
mathematically, a cheater with a powerful quantum computer can construct a
proper collective measurement corresponding to the mapping, and reads the
sealed bit with the strategy proposed in Ref. \cite{He} so that he can
escape the detection with a non-trivial probability. Hence the protocol is
easy to be broken. But in practice, we can make it more difficult to know
the rule of the mapping dishonestly than to do so honestly. One of such
methods is to translate the sentences describing the rules of the mapping
into a classical binary bit string, and seal it as a part of the sealed
string so that the cheater does not know the mapping before decoding the
string.

For example, Alice can first encode the following sentence into a classical
binary bit string
\begin{eqnarray*}
&&``\mathit{Measure\ the\ last\ two\ qubits\ in\ the\ basis\ } \\
&&\{\cos 15%
{{}^\circ}%
\left| 0\right\rangle +\sin 15%
{{}^\circ}%
\left| 1\right\rangle ,-\sin 15%
{{}^\circ}%
\left| 0\right\rangle +\cos 15%
{{}^\circ}%
\left| 1\right\rangle \}\mathit{\ } \\
&&\mathit{and\ you\ will\ know\ the\ value\ of\ the\ sealed\ bit\ from\ } \\
&&\mathit{their\ parity.\ Other\ qubits\ following\ this\ sentence\ } \\
&&\mathit{are\ all\ dummy\ qubits.\ You\ can\ simply\ leave\ them\ } \\
&&\mathit{alone.}"
\end{eqnarray*}
Then she seals it with our QSS protocol, and provides Bob the qubits
encoding this sentence, followed by a large number of qubits where only the
last two are actually useful.

We must point out that this bit sealing method is still insecure in
principle. This is because any given classical $N$-bit string can be decoded
into one specific sentence only, and the meaning of the sentence will lead
to a value of the sealed bit unambiguously. As long as a dishonest Bob knows
the length $N$\ of the sealed string, he can study all the $2^{N}$ possible
classical $N$-bit strings, decode them into sentences, and divide these
sentences into two subsets corresponding to the two possible value of the
sealed bit (of course there will also be tons of meaningless sentences and
even random sequences of bits that cannot be decoded as sentences. Bob can
simply leave them alone). Then as described in Ref. \cite{He}, he needs not
to know the content of the sealed sentence exactly. He simply constructs a
proper collective measurement to determine which subset the sentence belongs
to. Thus he will know the sealed bit from the $N$ qubits without disturbing
them seriously.

Nevertheless, if $N$ is sufficiently large, the number of possible sentences
will be enormous. There could be sentences as simple as
\[
``\mathit{It\ is\ }0\mathit{.\ Ignore\ the\ rest\ qubits.}"
\]
But there are also sentences like
\begin{eqnarray*}
&&``\mathit{Decode\ the\ bits\ following\ this\ sentence\ as\ a\ } \\
&&\mathit{bitmap\ image\ and\ count\ the\ parity\ of\ the\ } \\
&&\mathit{number\ of\ females.}"
\end{eqnarray*}
and followed by the bits encoding an image of animals, or famous
cartoon icons, or a picture as shown in Fig. 1. For an honest
participant who does not mind disturbing the quantum state sealing
the string, he can simply decode the whole string and learn the
value of the sealed bit easily. But for a dishonest participant who
wants to keep the quantum state less disturbed so that he can escape
the detection, he does not know beforehand what kind of the
sentences he will encounter. To decode the sealed bit with
collective measurements, he has to build a data base which contains
all sentences, pictures, and even wave files for voice messages,
etc., which can be encoded as $N$-bit strings. More importantly,
learning the sealed bit from pictures such as Fig. 1 is easy for
human beings with common senses, while if it is expected to be done
by machine, it has to have advanced power on image identifying,
pattern matching, and the capability to understand the meaning of
pictures, which may require a large coverage of the knowledge of
science and culture in human history. It is actually impossible in
practice for a cheater to build such a data base for sufficiently
large values of $N$, even he has the most powerful quantum computer.
Thus his collective measurements cannot be constructed. In this
sense, such QBS can be viewed as secure in practice.

\begin{figure}
\includegraphics{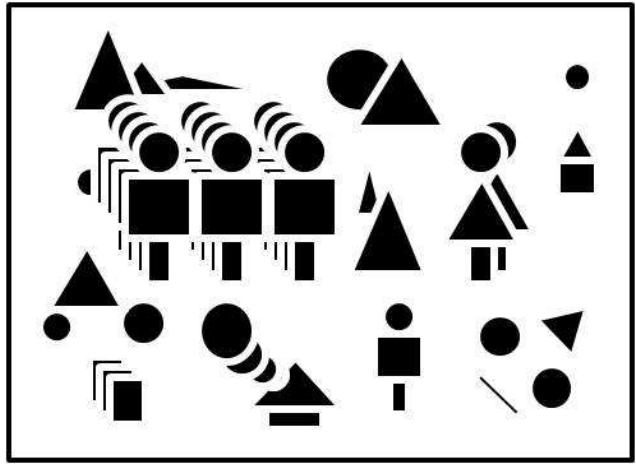}
\caption{\label{fig:epsart}A picture that may be used to seal a
single bit in practice.}
\end{figure}

With the above results of Ref. \cite{String}, building a QBC protocol is
straight forward. We can simply apply this ``practically secure'' QBS
protocol in Advanced Protocol to obtain a QBC protocol. Note that the
``practically secure'' QBS does not satisfy the strong unconditional
security condition. In fact, it is not unconditionally secure in principle
at all. Consequently, the resultant QBC protocol is not unconditionally
secure in principle. But since the above QBS cannot be broken in practice,
the resultant QBC, whose security completely determined by the QBS protocol
being based on, is also secure in practice. Especially, when $N$ is
sufficiently large, there will inevitably some sequences of bits which can
be decoded incidently as sentence like
\begin{eqnarray*}
&&``\mathit{Go\ to\ the\ main\ library.\ Find\ the\ book\ on\ the\ } \\
&&\mathit{top-left\ of\ the\ last\ shelf.\ Turn\ to\ the\ last\ page,\ } \\
&&\mathit{and\ count\ how\ many\ times\ the\ letter\ K\ occurs\ } \\
&&\mathit{in\ the\ 3rd\ line......}"
\end{eqnarray*}
Though such a sentence may lead to a deterministic value of the
sealed bits, it is less likely that the bit is really sealed this
way according to common sense. Therefore a cheater can simply keep
such sentences out of consideration in a QBS protocol. Nevertheless,
in our QBC protocol we can allow Bob (again, please note that the
names Alice and Bob called in Advanced Protocol are contrary to
those in QBS) to seal such ridiculous sentences (and even
meaningless sentences which is literally correct but may contain no
clue for the sealed bit) with a certain probability which is also
agreed by Alice. Whenever Alice decodes such a sentence in step
(iii), she can tell Bob to discard the corresponding data and
chooses another quantum register to measure. This will make the
protocol even more secure because if Alice wants to cheat, her
quantum computer will also have to deal with all ridiculous and
meaningless sentences, and judge which sentences make sense. To
accomplish this, it seems that the quantum computer needs not only
unlimited computational power to perform any kind of quantum
operation and measurement, but also some extent of artificial
intelligence. So far there is even debate on whether human
intelligence is computable or not \cite {Penrose}. Therefore the
security of our protocol is further guaranteed by those incomputable
elements. As a result, though the MLC no-go theorem is considered to
have put a serious drawback on the development of the theoretical
aspect of quantum cryptography, in practice we can simply bypass
this theorem and go ahead with this ``practically secure'' QBC
protocol. After all, the main purpose of developing cryptography is
to satisfy the application needs of human society, which involve not
only physical and mathematical principles, but also human behaviors.
Cheating itself is exactly a creation of human intervention.
Therefore it is natural to introduce more human-related issues into
cryptography to prevent cheating. Our finding makes it possible to
building any complicated QBC-based ``post-cold-war era'' multi-party
cryptographic protocols, like quantum coin tossing and quantum
oblivious transfer, and achieve better security in practice which
cannot be reached by their classical counterparts.

\section{Feasibility}

The present QBC protocol can be implemented as long as the based-on
QS protocol can be realized. As pointed out in Ref. \cite{String},
the QS protocol requires merely Alice to have the ability to prepare
each single qubit in a pure state, and Bob to perform individual
measurements. No entanglement/collective-measurement is required.
Therefore the protocol may be demonstrated and verified by the
techniques available currently. Of course for practical uses,
storing quantum states for a long period of time is still a
technical challenge today. But this is a tough issue that all QS
protocols have to suffer. The present protocol may be one of the
simplest protocols of its kind. More significantly, if such
technique of storing quantum states is not available, the attack
strategy proposed by the MLC no-go theorem that makes previous QBC
(e.g, Ref. \cite{BCJL93}) insecure cannot be implemented either; but
on the other hand, whenever this tough technical problem is resolved
in the future, the previous QBC would be broken quite easily, while
our QBC protocol may still remain to be secure in practice.

The work was supported by the NSFC under grant Nos.10605041 and 10429401,
the RGC grant of Hong Kong, the NSF of Guangdong province under Grant
No.06023145, and the Foundation of Zhongshan University Advanced Research
Center.


\begin{thebibliography}{99}
\bibitem{BCJL93}  G. Brassard, C. Crepeau, R. Jozsa, and D. Langlois, in
\textit{Proceedings of the 34th Annual IEEE Symposium on Foundations
of Computer Science }(IEEE, Los Alamitos, 1993), p.362.

\bibitem{Kilian}  J. Kilian, in \textit{Proceedings of 1988 ACM Annual
Symposium on Theory of Computing} (ACM, New York, 1988), p.20.

\bibitem{Yao}  A. C. C. Yao, in \textit{Proceedings of the 26th Symposium on
the Theory of Computing} (ACM, New York, 1995), p.67; G. P. He and
Z. D. Wang, \textit{Phys. Rev.} \textbf{A 73}, 012331 (2006); {\it
ibid.} \textbf{A 73}, 044304 (2006).

\bibitem{Mayers}  D. Mayers, \textit{Phys. Rev. Lett.} \textbf{78} (1997)
3414.

\bibitem{LC}  H. -K. Lo and H. F. Chau, \textit{Phys. Rev. Lett.}\textbf{78}
(1997) 3410.

\bibitem{sealing}  H. Bechmann-Pasquinucci, \textit{Int. J. Quant. Inform.}
\textbf{1}, 217 (2003).

\bibitem{impossibility}  H. Bechmann-Pasquinucci, G. M. D'Ariano, C.
Macchiavello, \textit{Int. J. Quant. Inform.} \textbf{3,} 435 (2005).

\bibitem{He}  G. P. He, \textit{Phys. Rev.} \textbf{A 71} (2005) 054304.

\bibitem{String}  G. P. He, \textit{Int. J. Quant. Inform.} \textbf{4}, 677
(2006).

\bibitem{Insecure}  H. F. Chau, \textit{Phys. Rev.} \textbf{A} \textbf{75},
012327 (2007).

\bibitem{note} Although it was claimed in Ref.~\cite{Insecure} that all
QSS are insecure, as indicated later in Ref.~\cite{Security}, the
cheating strategy proposed in Ref. \cite{Insecure} is not a
successful cheating because it cannot obtain non-trivial amount of
information while escaping the detection simultaneously
\cite{Security,qi505}. It was also realized in Ref. \cite{qi516}
that when the cheating in Ref. \cite{Insecure} escapes the
detection, the ratio between the amount of information obtained by
the cheater and that of the sealed string is arbitrarily small as
the length of the string increases.

\bibitem{Security}  G. P. He, \textit{Phys. Rev.} \textbf{A 76}, 056301
(2007); quant-ph/0602159 v1.

\bibitem{qi505}  M. Nakanishi, S. Tani, S. Yamashita, in \textit{Proc. of
the 6th WSEAS International Conference on Information Security and Privacy
(ISP'07)} (WSEAS, Spain, 2007), p.30.

\bibitem{qi516}  H. F. Chau, \textit{Phys. Rev.} \textbf{A 76}, 056302
(2007).

\bibitem{Penrose}  R. Penrose, \textit{Shadows of the Mind: A Search for the
Missing Science of Consciousness}, Oxford 1994; \textit{The Large, the Small
and the Human Mind}, Cambridge 2000.

\bibitem{code}  F. J. MacWilliams and N. J. A. Sloane, \textit{The Theory of
Error-Correcting Codes} (North-Holland, Amsterdam, 1977).
\end{thebibliography}
\end{document}